# Nano tracks in fullerene film by dense electronic excitations

P. Kumar[1], D. K. Avasthi[2], J. Ghatak[3], P. V. Satyam[3], R. Prakash[1], A. Kumar[1*]

[1]School of Material Science and Technology, Indian Institute of Technology (BHU), Varanasi - 221005, India

[2]Inter-University Accelerator Centre, PB- 10502, New Delhi 110067, India

[3]Institute of Physics, Sachivalaya Marg, Bhubaneswar -751 005, India

**Abstract**

In the present work, we investigate the formation of nano tracks by cluster and mono-atomic ion beams in the fullerene ($C_{60}$) thin films by High Resolution Transmission Electron Microscopy (HRTEM). The fullerene films on carbon coated grids were irradiated by 30 MeV $C_{60}$ cluster beam and 120 MeV Au mono-atomic beams at normal and grazing angle to the incident ion beams. The studies show that the cluster beam creates latent tracks of an average diameter of around 20 nm. The formation of large size nano tracks by cluster beam is attributed to the deposition of large electronic energy density as compared to mono-atomic beams.

**\*Corresponding author:**
　　　　　Email : amit.mst@iitbhu.ac.in,　& pawan.iit13@gmail.com

# Introduction

Fullerenes are important materials for the fabrication of nano-devices such as field effect transistors, flat panel display based on field emission [1, 2]. The controlled conducting nano wire patterns formation in films by ion beam has been demonstrated [3, 4]. Phase transformation of fullerene by ion beam has been a topic of fundamental research interests. Ion irradiation induced phase transformation is one of the most important phenomena in the field of ion beam modification of fullerenes [5, 6]. It is known that energetic heavy ions with velocities comparable or higher than the Bohr velocity provide a unique tool to create the extremely localized conditions of temperature and pressure in nanometric volumes for an ultra-short time, typically of the order of picoseconds. Each ion induces a continuous trail of damage having a few nanometer diameter and typically several tens of micrometers length. There have been several studies in the past, including a few from our group of irradiation effects of high energy heavy ions in fullerene films [7-10]. Special feature of high energy heavy ion is that the energy lost by the ion (that brings the changes in the film under ion irradiation) is dominantly via inelastic collisions, in contrast to low energy ions where the energy loss is dominantly via elastic collisions. The existence of an electronic stopping power threshold ($dE/dx$) for the phase transformation is well the established for different types of materials (dielectric, semiconductors and metals) [11-13]. Vast experimental data are now available concerning the dependence of induced disorder on ($dE/dx$) but the damage mechanism is still an open question. Various mechanisms have been proposed to explain the transfer of energy from the energy deposited by ion to the atomic motion in the material, two main models are the thermal spike [14] and the Coulomb explosion [15].

Nowadays, the availability of cluster ion beams in the MeV range extends the possibilities of increasing the density of energy deposition [16, 17]. As the cluster ions pass through the solids, they break up within the first few atomic layers and the constituent atoms deposit their energy simultaneously. The trajectories of atoms (of fragmented cluster) remain strongly correlated over a distance that depends on the initial energy and number of constituents [18]. After a certain distance, strong correlation of trajectories is lost and hence the density of electronic energy deposition reduces drastically. Recently, the formation of latent tracks in semiconductors (Si, Ge, GaAs) by cluster beams has been reported at room temperature [19, 20], whereas tracks are not created by mono-atomic beams in these materials. In the present study we investigate dense electronic energy deposition effects in fullerene films, which have semiconducting properties with optical band gap ~2 eV. In this material, there have been a few studies of the damage induced by swift heavy ions, using FTIR, Raman and other techniques [7, 10, 21, 22]. The formation of tracks in fullerene films was first explained by Dufour et al. [23], in the electronic energy loss region from 3 keV/nm to 10 keV/nm by thermal spike calculations. They reported that tracks should be formed above a threshold $(dE/dx)_e$ value of about 3-4 keV/nm. Although there have been a number of studies on damage of fullerene by ion beam [3,4,7-10] indicating the possible formation of ion tracks, no direct evidence of tracks in fullerenes by microscopy techniques has been reported. The damage process under the dense electronic excitations in such a complicated systems needs to be investigated in more details. In the present work, we study the formation of tracks in fullerene films by high resolution transmission electron microscopy. For this purpose, we irradiated fullerene films on carbon coated grids by 30 MeV $C_{60}$ and 120 MeV Au ion beams.

**Experimental Details**

Fullerene thin films were deposited on carbon coated transmission electron microscopy (TEM) grids in a vacuum of $1\times10^{-6}$ Torr by resistive heating using commercially available 98% pure $C_{60}$ in a tantalum boat. The thickness of the film as measured by quartz crystal thickness monitor was about 30 nm. The samples were irradiated with 30 MeV $C_{60}$ cluster and 120 MeV Au mono-atomic ion beams from Tandem accelerator at IPN Orsay, France and 15 UD Pelletron at IUAC New Delhi, India, respectively. Samples were irradiated with $C_{60}$ cluster ions at normal and grazing incidence up to fluences of $10^{10}$ ions cm$^{-2}$. This range of fluences was chosen in order to avoid any spatial overlap of tracks. In the case of mono-atomic ions, TEM samples were irradiated with fluences up to $2\times10^{11}$ ions cm$^{-2}$. All irradiations and TEM measurements were performed at room temperature. The electronic energy loss and nuclear energy loss were estimated by a computer code Stopping and Range of Ion in Matter (SRIM-2013) [24]. The energy loss per incident cluster is the sum of the energy loss of individual carbon atoms as experimentally checked by energy loss measurements for various cluster ions [25]. The electronic energy loss by 30 MeV $C_{60}$ and 120 MeV Au ion beam were ~ 40.2 keV/nm and ~13.7 keV/nm, respectively. Similarly, nuclear energy loss by 30 MeV $C_{60}$ and 120 MeV Au ion beam were ~ 25 eV/nm and ~ 162 eV/nm, respectively. The samples irradiated at grazing incidence (about 80° with respect to normal from surface) have the advantage that the evolution of the track shape during the slowing-down of the projectile can be studied by TEM observations. The samples were analyzed using a JEOL 2010 UHR transmission electron microscope (TEM) operated at 200 kV, at Institute of physics, Bhubaneswar, India. The used high resolution transmission electron microscopy has 0.19 nm point to point resolution. The electron irradiation during TEM measurement was observed to be very destructive for fullerene [26, 27]. To avoid any modification in the fullerene matrix by electron beam during TEM measurement, the electron flux was kept as low as possible and the electron beam spot was shifted periodically during acquisition.

**Results and discussion**

The nano track evolution in fullerene matrix by cluster ion ($C_{60}$) is induced in the fullerene films irradiated up at normal and grazing angle to a fluence of $1\times10^{10}$ ions.cm$^{-2}$. Figure 1 shows a bright field (BF) image of latent tracks formation in fullerene irradiated by 30 MeV $C_{60}$ ions at normal incidence at room temperature. The image permits to determine the density of the tracks as well as their diameter. The tracks are randomly distributed all over the surface of the sample and the track density is consistent with the fluence of projectiles. Figure 2 depicts a high resolution TEM (HRTEM) images of the pristine (Figure 2a) and irradiated fullerene films at normal incidence (Figure 2b). The lattice fringes of pristine $C_{60}$ film are seen in Figure 2a, with spacing d ~ 0.82 nm, which matches with reported values [22]. An impact of one cluster ion is observed in Figure 2b. From the HRTEM image, the impact region appears to be amorphous and non-impact region to be crystalline in the nature. Low magnification image (shown in Figure 1 and Figure 3) were taken under the conditions of BF imaging (low magnification BF images that are usually imaged with forward beam - by using a single beam). The high resolution images were taken by using few beams (at least two beams) under very high magnification, and give the lattice fringes due to the phase contrast in the scattered electron waves. It is interesting to note that Fresnel rings appear at the low magnifications also arise due to phase contrast but happens because of existence of a sharp boundary or edge or a hole. These Fresnel fringes from any edge in an out-of-focus electron image (over focus or under focus). These Fresnel fringes can be used to confirm the existence of an interface, such as a track formation in the present research work. These can be done under bright field image conditions as well (i.e. imaging with single forward beam) as the interfacial edge helps to generate another secondary wave. It is known that these fringes have been used in correcting the lens astigmatism in a TEM [28].

The tracks are seen clearly in the Figure 3, which depicts TEM micrographs taken with (a) focus condition (b) over-focused and (c) under-focused. The Fresnel fringes that appear in Figure 3(b) (with a white fringe around the track) and weakly in Figure 3(c) (with a white fringe around the track). The TEM images reveal the average track diameter in the around 20 nm. The sharp boundary, because of which fringes can be seen, lies at the interface of the amorphous region and crystalline region in and around the track.

To investigate the track shape in the depth of fullerene matrix, we performed some irradiation at grazing incidence (about 80° with respect to normal from surface). Figure 4 shows the shape of tracks in a sample irradiated at tilt angle of 80°, the arrow shows the direction of cluster beam. We see that the diameter of track remains constant up to some depth and that the track diameter shrinks deeper in the material until the track contrast disappears. The shape of the track is due to the fact that the cluster fragments into smaller size constituents in first few atomic layers and that the fragments move in correlation among themselves to a 2% longer in-depth than the lateral track size. The end of the track region represents the location where the number of correlated constituents is so low that the addition of their effects does not permit the reach the threshold of energy deposition for track registration.

HRTEM imaging on 120 MeV Au mono-atomic ions at fluence of $2\times10^{11}$ ions.cm$^{-2}$ irradiated fullerene film do not reveal any well-defined tracks. However, we observed that the lattice is more or less homogeneously damaged on same area due to energetic ion impacts. The formation of track in the fullerene films has been reported [7, 22] by conducting atomic force microscopy. The authors have also evaluated the effective radii of tracks of energetic ions in irradiated fullerene films from damage cross-section using Raman and FTIR spectroscopy techniques [29-34]. However there has not been any direct microscopic experimental evidence of ion track formation in fullerene by mono atomic ion beams.

To get a comprehensive picture of ion track formation in fullerene thin film by the dense electronic energy deposition, a comparison of energy losses and effective track radius is documented in the table 1, for a wide range of electronic energy loss $S_e$ values from 4 to 4020 eV/Å. The damage cross-sections ($\sigma$) have been estimated using Raman-spectroscopy and FTIR analysis for various ion (H, He, Ar, O, Fe, Ni, Ag, Au) irradiated fullerene film [29-36]. The track radius ($r_e$) for various ions are calculated using the $\sigma = \pi (r_e)^2$ relation. In the present HRTEM studies revealed the direct observation of nano-size track formation using $C_{60}$ cluster ion in the fullerene film.

Previous reports on the effective track formation in the fullerene are mainly based on indirect methods (estimation of effective radius from the damage cross-sections, $\sigma$) by monitoring the distractions of fullerene molecules in the films. Figure 5 shows the variation of track radius ($r_e$) as a function of electronic energy deposition ($S_e$). It is observed that the track radius linearly increases with the higher energy deposition ($S_e$). Up to 258 eV/Å energy deposition there is no significant track formation, which starts to increase significantly for 70 MeV Fe [30, 31] and 110 MeV Ni [29] ions (higher electronic energy) onward. On irradiation of $C_{60}$ cluster ions there is huge electronic energy deposition in the fullerene film in a confined ion impact zone which results in formation of recognized track of ~ 10 nm radius as observed in high resolution TEM images in figure 2. TEM imaging is a direct microscopic method to get the exact track diameter and clearly revealed the nano-track formation in fullerene by 30 MeV $C_{60}$ cluster ion irradiation. It is noticed that the track radius almost linearly increases with the increase of electronic energy deposition after certain threshold value i.e. size of track formation is directly proportional to the electronic energy loss, $S_e$ {$\sigma \propto (dE/dX)^2$} which follows the coulomb explosion model [15]. It also support the previous report [33] for different ions (He, C, O, S, Br, I) of varying $S_e$ values on $C_{60}$ film.

We have also tried to fit our experimental data considering the thermal spike model calculations [23]. Thermal spike model has been used to understand the track formation in metallic as well as insulator materials, where track radius mainly depends upon the electron phonon coupling (g) and energy (ΔH) require melting the materials [23]. It is noticed that track radius for low electronic energy losses show a better agreement with the reported thermal spike calculations for fullerene thin films but significantly deviate at much higher energy loss values. Similar deviations at higher energy loss were also reported on bismuth targets [23]. Therefore thermal spike model required more refinement to understand the track formation at much higher electronic energy loss. Linear increase of track radius with electronic energy loss in our analysis shows that the track formation in fullerene thin films can be qualitatively understood using Coulomb explosion model [15]. It is interesting to notice that for lower $S_e$ values, there is validation of synergetic effect of $S_e$ and $S_n$ in damage creation as clearly seen in the inset of figure 5. In lower $S_e$ regimes there is no significant variation in $r_e$, but for few cases having larger values of $S_n$ (e.g. Ar ions) there is distinct increment in track size. The synergetic effect of $S_e$ and $S_n$ has already been observed in the ion irradiation process of various materials [37, 38] and ion beam induced crystallization [39-41].

The energy per unit volume is more relevant parameter to understand the track formation and to compare the diameters of tracks obtained with cluster beams and mono-atomic ion beams rather than the linear rate of electronic energy loss [19]. It is noticed that to calculate the energy deposition per unit volume, the energy or range of the secondary electrons is required. These electrons play crucial role in the radial extension of the deposited energy, which is much larger in case of energetic ion beam irradiation [42]. The energy of the secondary electrons can be calculated from the maximum energy transferred to the electrons ($E_m = 2m_e v^2$; $m_e$ is the electron mass and v is the projectile ion velocity) and using a semi-empirical range-energy relationship [43]. In the case of cluster ions, the energy of delta electron is around 92 eV and their range is few nm, but in case of mono-atomic ion (120 MeV Au) the energy of the delta electron is few keV and their corresponding range is about few hundred nm. The electronic energy density deposition qualitatively can be extracted by a simple relation $E_d = [1/\pi r^2 \cdot (dE/dx)_e]$, where r is the average diameter of latent tracks, and the $(dE/dx)_e$ is electronic energy loss. Therefore energy density deposited by cluster is about four orders of magnitude higher that of mono-atomic ions (120 MeV Au). At low velocities (in cluster ion irradiation), the secondary electrons produced by the passage of the cluster ion have smaller range therefore the deposited energy density is higher. Whereas in the case of high energy ion (120 MeV Au), the range of delta electron is few hundreds of nm, therefore the energy density deposition is smaller, which is not sufficient for observable track formation. For the systems under study, it is possible to determine the track formation with details about possible track diameter. As the present data is a projection data, this limits any 3-dimensional view of tracks. The contrast is basically arising due to crystalline and amorphous regions.

However, a defective region along the path of the projectiles have been deduced from Raman/FTIR spectroscopy and conducting AFM measurements. Present observations as well as analysis will be useful to develop the new theoretical models to get detail understanding of nano-track formation in fullerene matrix and energetic ion interaction with molecular materials. Using such ion interactions, it is possible to controllably manipulate the material properties in various ways and useful to create the confined nano scale structures. Dense electronic excitations by the energetic ion beams may enable many important applications in nanoscale engineering such as the production of regularly ordered arrays of nanowire, quantum dots (QDs) etc., which may be useful for efficient field emitters, optical emitters and large scale technological applications in future.

## Conclusion

Present work reports the investigation of ion track formation in fullerene thin film created by the cluster ($C_{60}$) and mono-atomic ion beams. HRTEM measurements revealed that the cluster beam lead to formation of recognizable latent track (radius ~10 nm) in the fullerene thin films. In case of mono-atomic Au ion, no tracks were observed under high resolution TEM imaging. Different electronic energy density deposition in mono-atomic and cluster beams is a crucial parameter to understand the track formation in fullerene films, which is mainly governed by the delta electrons produced during the ion irradiation. A comprehensive picture is provided to understand the track formation in fullerene matrix and variation of the track radius as a function of electronic energy loss.

## Acknowledgments

We gratefully acknowledge the financial support from the Department of science and technology (DST), New Delhi, India. We would like to thank Dr. A. Dunlop, Laboratoire des Solides Irradiés, Eco., Orsay (France) for providing the $C_{60}$ cluster ions facilites. We are also thankful to both the accelerator groups at IUAC, New Delhi and IPN Orsay institutes for providing the stable ion beams.

**TABLE 1.** Track radius estimated from damage cross-section ($\sigma$) for various energetic incident ions having different $S_e$ and $S_n$ values calculated from TRIM code.

| Ion Energy | $S_e$ (eV/Å ion) | $S_n$ (eV/Å ion) | Damage Cross section, $\sigma$ for $C_{60}$ (cm$^2$) | Track Radius, $r_e$ (nm) | Ref. |
|---|---|---|---|---|---|
| 30 MeV $C_{60}$ | 4020.0 | 2.50 | *$3.1 \times 10^{-12}$ | 10.0 | Present work |
| 100 MeV Au | 1274.0 | 1.90 | $5.0 \times 10^{-13}$ | 4.0 | 29 |
| 120 MeV Ag | 1000.0 | 3.60 | $4.0 \times 10^{-13}$ | 3.5 | 32 |
| 78 MeV I | 942.0 | 7.65 | $1.9 \times 10^{-13}$ | 2.5 | 33 |
| 48.6 MeV Br | 716.0 | 3.60 | $1.1 \times 10^{-13}$ | 1.9 | 33 |
| 110 MeV Ni | 673.0 | 0.98 | $8.5 \times 10^{-14}$ | 1.6 | 29 |
| 70 MeV Fe | 654.0 | 1.22 | $4.5 \times 10^{-14}$ | 1.2 | 30,31 |
| 19.7 MeV S | 375.0 | 0.84 | $3.2 \times 10^{-14}$ | 1.0 | 33 |
| 980 MeV Fe | 258.0 | 0.12 | $1.1 \times 10^{-14}$ | 0.6 | 30,31 |
| 70 MeV O | 79.6 | 0.04 | $1.4 \times 10^{-15}$ | 0.2 | 29 |
| 0.22 MeV Ar | 64.6 | 32.70 | $3.9 \times 10^{-14}$ | 1.1 | 34,35 |
| 0.3 MeV C | 54.5 | 1.87 | $2.1 \times 10^{-15}$ | 0.3 | 34,35 |
| 0.3 MeV He | 30.8 | 0.09 | $2.2 \times 10^{-16}$ | 0.08 | 34,35 |
| 0.1 MeV H | 12.7 | 0.02 | $6.1 \times 10^{-17}$ | 0.04 | 34,35 |
| 0.16 MeV H | 11.2 | 0.01 | $5.0 \times 10^{-17}$ | 0.04 | 34,35 |
| 0.002 MeV Ar | 7.0 | 46.00 | $8.5 \times 10^{-15}$ | 0.5 | 34,35 |
| 1 MeV H | 4.0 | 0.002 | $2.0 \times 10^{-17}$ | 0.03 | 36 |

* $\sigma$ being estimated by the measured track radius.

**Figure captions**

Fig. 1. Transmission electron micrograph of latent tracks formed in fullerene film under 30 MeV $C_{60}$ cluster beam irradiation. The direction of beam was normal to the sample surface.

Fig. 2. HREM images of (a) pristine fullerene matrix (b) latent track formation under 30 MeV $C_{60}$ cluster beam irradiation.

Fig. 3. Transmission electron micrograph of 30 MeV cluster ion impact fullerene film at normal to ion beam (a) Bright field image in focus condition (b) Phase contrast in over-focused and (c) Phase contrast in under-focused.

Fig. 4. Transmission electron micrograph of fullerene film irradiated with 30 MeV cluster ion at about 80 degree incidence.

Fig. 5. Variation of Track radius ($r_e$) with increasing $S_e$ values for various energetic ions linearly fitted with 8% error bar. Inset showing logarithmic scale of $S_e$ values for the distinguished lower and higher value effects.

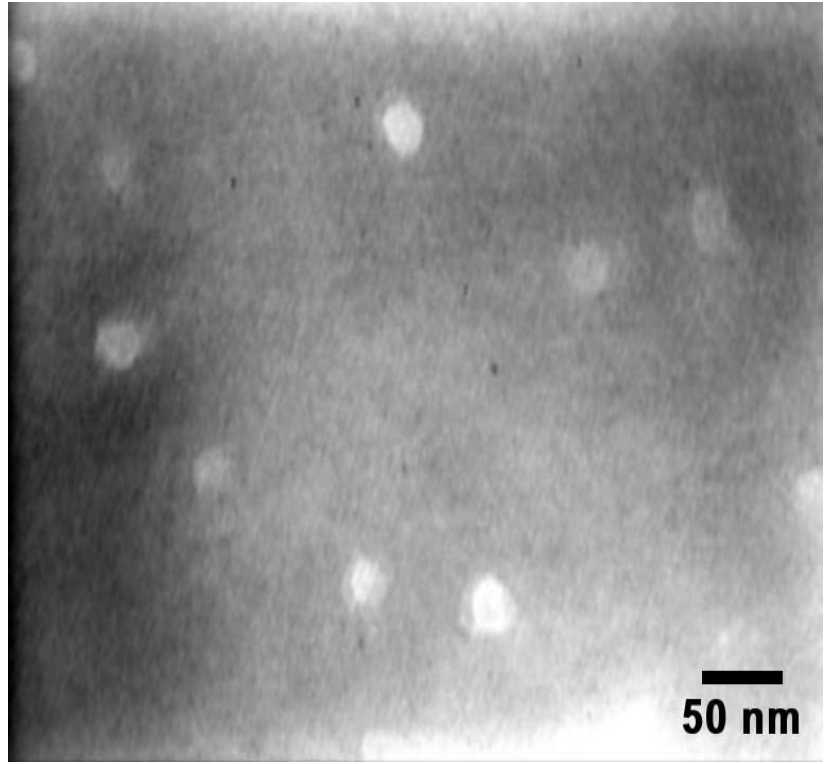

**Figure 1**

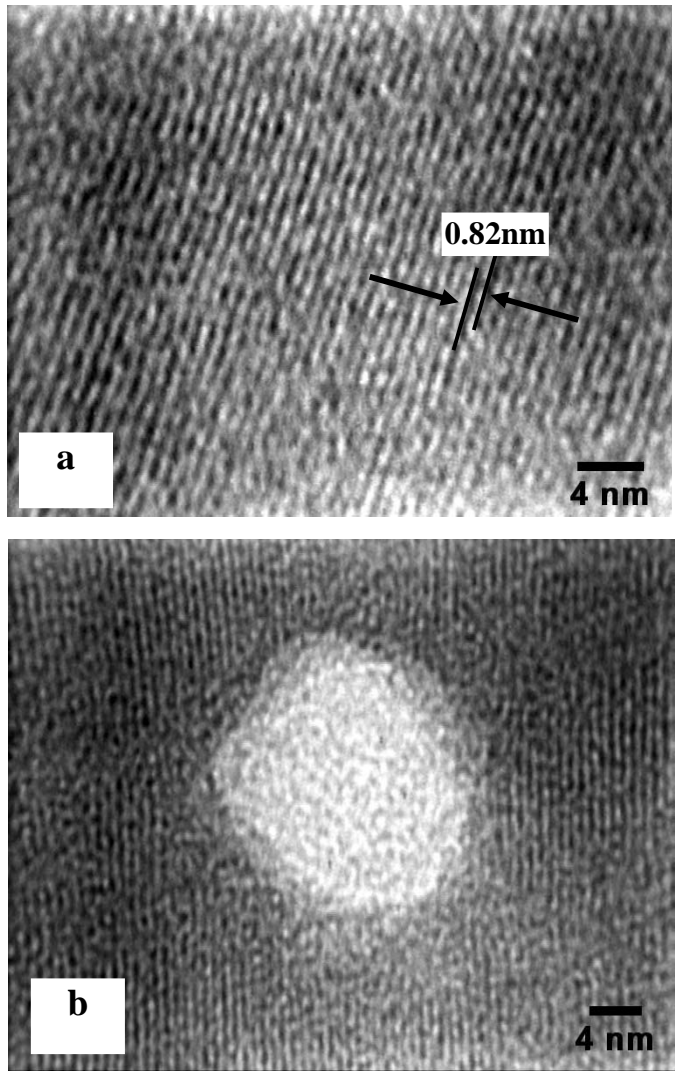

**Figure 2**

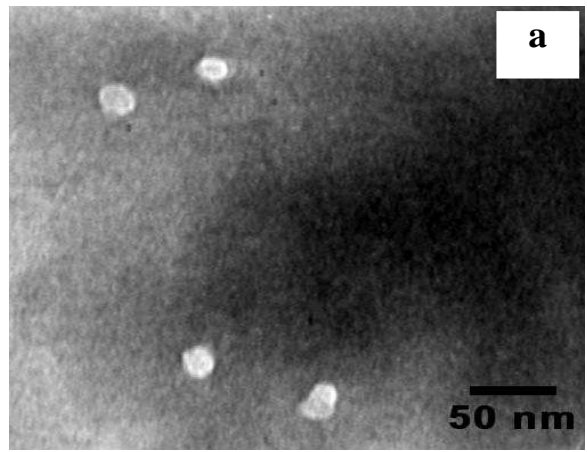
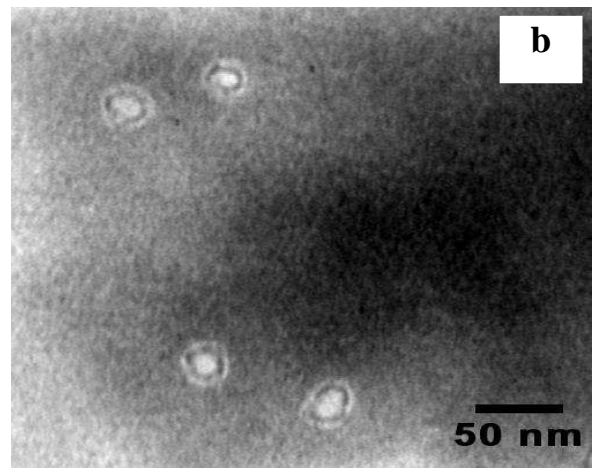
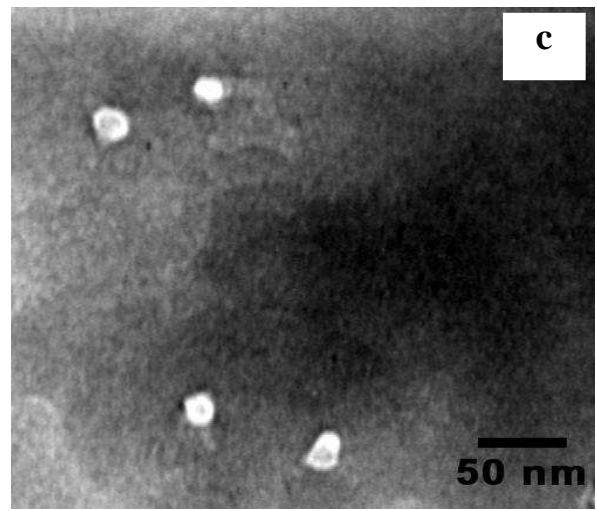

**Figure 3**

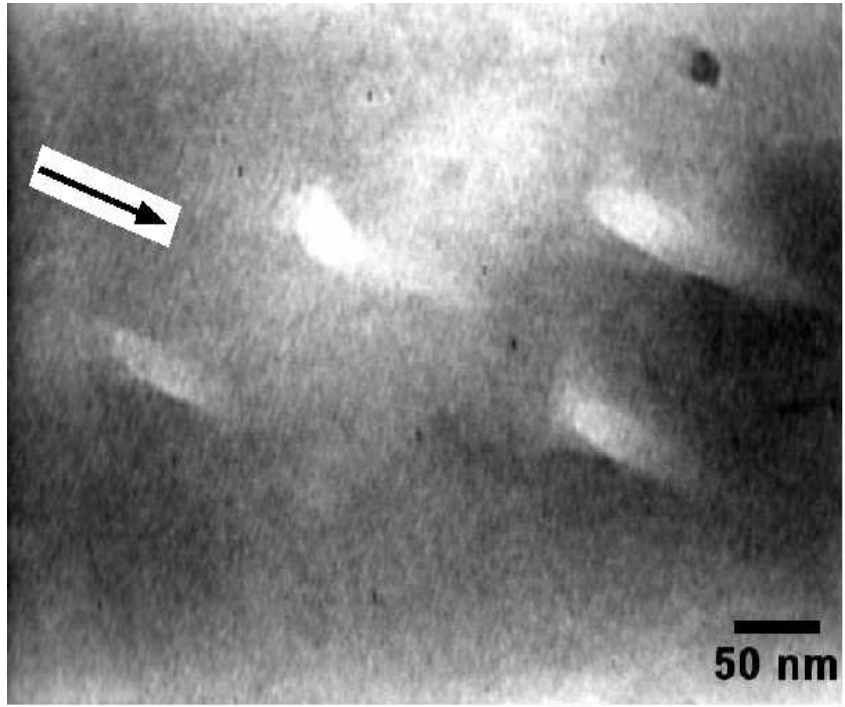

**Figure 4**

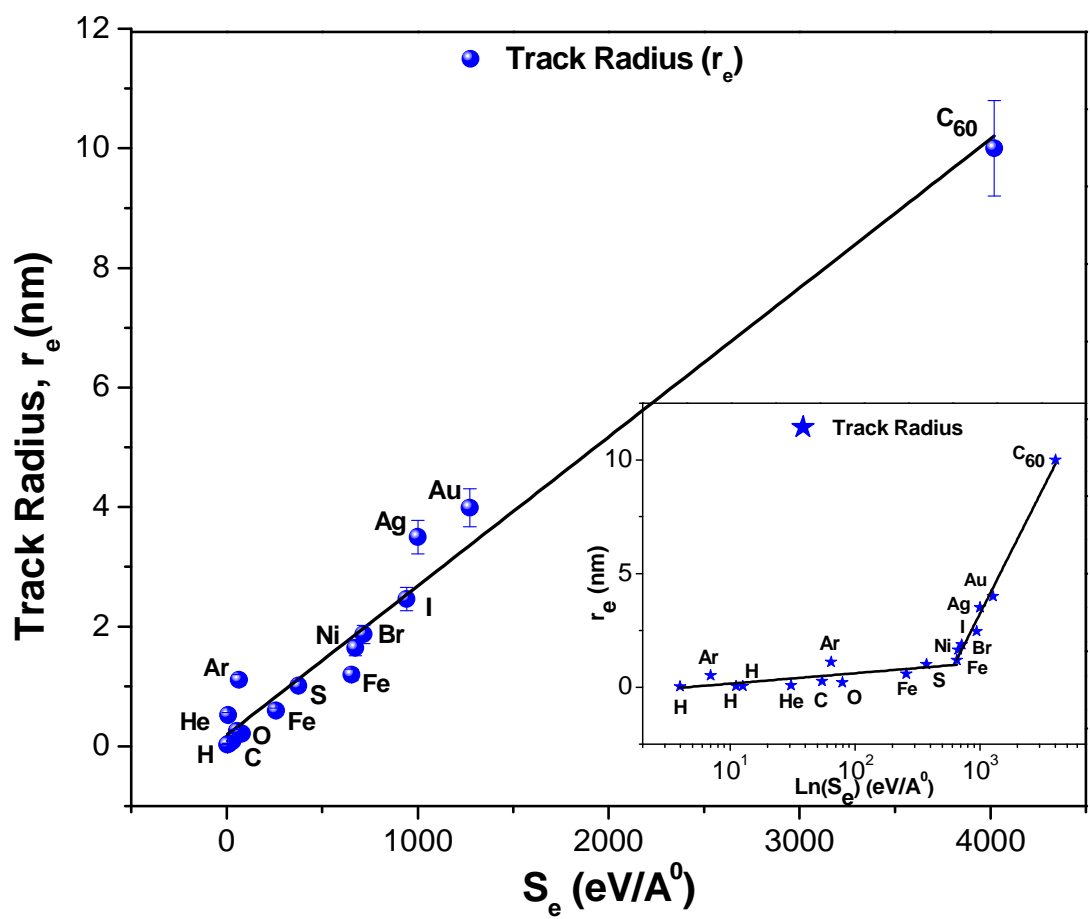

**Figure 5**